\documentclass{PoS}
\usepackage{amsmath,amssymb,bm}
\usepackage{graphicx}
\usepackage{subfigure}
\usepackage{xspace}
\usepackage{accents}
\usepackage{units}
\usepackage{booktabs}
\usepackage[above]{placeins}
\usepackage{colortbl}
\usepackage{xcolor}
\usepackage{verbatim}
\usepackage{upgreek}

\definecolor{darkred}{rgb}{0.5,0,0}
\definecolor{darkblue}{rgb}{0,0,0.5}
\definecolor{firebrick}{rgb}{0.75,0.125,0.125}
\definecolor{darkgreen}{rgb}{0,0.5,0}
\definecolor{darkyellow}{rgb}{0.5,0.5,0}
\definecolor{darkcyan}{rgb}{0,0.5,0.5}
\definecolor{rulecolor}{gray}{0.8}
\usepackage{multirow}
\def\Geant4{{\sc Geant4}}

\graphicspath{{./}}

\title{High-energy interactions \\ at the Pierre Auger Observatory}

\ShortTitle{High-energy interactions at the Pierre Auger Observatory}

\author{\speaker{Ruben Concei\c{c}\~ao}$^a$\thanks{R. Concei\c{c}\~ao gratefully acknowledges the financial support by FCT (SFRH/BPD/73270/2010)}\ \ for the Pierre Auger Collaboration$^b$\\
	\llap{$^a$}LIP / IST, Lisboa, Portugal\\
	\llap{$^b$}Observatorio Pierre Auger, Av. San Mart\'in Norte 304, 5613 Malarg\"ue, Argentina\\
	E-mail: \href{mailto:auger_spokespersons@fnal.gov}{\rm
		auger\_spokespersons@fnal.gov}\\
	Full author list:
	\href{http://www.auger.org/archive/authors_2015_06.html}{\rm
		http://www.auger.org/archive/authors\_2015\_06.html}}

\abstract{The interaction of Ultra High Energy Cosmic Rays (UHECRs) with the atoms of the atmosphere can occur at center-of-mass energies that surpass 100 TeV, while present human-made accelerators go up to 13 TeV. Therefore it provides a unique opportunity to explore hadronic interactions at the highest energies. However, the extraction of hadronic interaction properties from the Extensive Air Showers (EAS) characteristics, which are induced by the UHECR, is intrinsically related to the nature of the primary cosmic ray. As such, to break the degeneracy between hadronic interactions and primary mass composition, a consistent description of the shower observables must be achieved. 

Such detailed studies have been conducted in the last years at the Pierre Auger Observatory, the largest UHECRs detector in the world. It combines two complementary techniques to measure the EAS characteristics.

In this talk, we will present the latest measurements on shower observables, both on the electromagnetic and muonic shower components, and its interpretation in terms of the primary mass composition. Its impact regarding particle physics will be discussed, in particular the measurement of the proton-air cross section.

Finally, through the joint analysis of the different measurements, it will be shown that none of the post-LHC high-energy hadronic interaction models can satisfactorily describe the data.}

\FullConference{The European Physical Society Conference on High Energy Physics\\
		22--29 July 2015\\
		Vienna, Austria}

\begin{document}

\section{Introduction}

Ultra High Energy Cosmic Rays are the most energetic known particles and yet their origin and nature remain a mystery. These particles arrive at Earth with a very scarce flux making its direct detection practically impossible. Fortunately, the interaction of UHECRs with the Earth atoms at the top of the atmosphere produces huge particle showers that can be detected. These are known as Extensive Air Showers (EAS), and can be either sampled at ground or, in moonless nights, can be detected via the collection of the fluorescence photons that are produced by the atmosphere nitrogen molecules due to the development of the shower.

The Pierre Auger Observatory~\cite{PierreAuger} is the world largest experiment to study UHECRs. It has a surface detector (SD) composed by more than 1600 Water Cherenkov Detectors (WCD), over an area of $3000\,{\rm km^2}$, to sample the charged secondary shower particles that reach the ground. It comprises also 4 Fluorescence Detector (FD) buildings, each with 6 telescopes, that survey the whole array.

The hybrid nature of the Pierre Auger Observatory (SD+FD) is one of its strongest features. An absolute energy scale can be derived for the SD, whose energy reconstruction is model dependent, using high quality shower events recorded independently by each technique. Also the geometrical shower reconstruction benefits from the joint analysis. Moreover, hybrid event analyses provide the observatory with an additional handle on the understanding of EAS physics.

\section{UHECR energy spectrum}

\begin{figure}[htbp]
	\centering %
	\begin{minipage}[c]{0.65\textwidth}%
		\centering \includegraphics[width=0.99\textwidth]{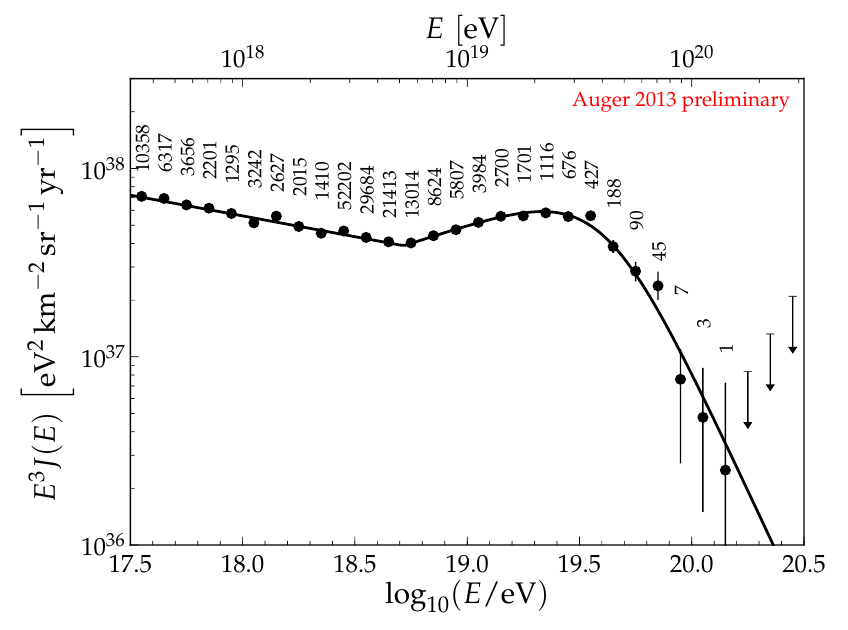} %
	\end{minipage}
	\protect\caption{The combined energy spectrum of UHECRs as measured at the Pierre Auger Observatory. The numbers give the total number of events inside each bin. The last three arrows represent upper limits at 84\% C.L.~\cite{spectrum}}
\label{fig:spectrum}
\end{figure}

The cosmic ray energy spectrum encloses information about the UHECRs sources and its distribution, production mechanisms and propagation. The results of the measurement of the UHECRs arrival flux from a combined analysis of SD (vertical and inclined events) and hybrid events (SD + FD)~\cite{spectrum} are presented in figure~\ref{fig:spectrum}. This energy spectrum has been measured using about 8 years of data\footnote{an update of this measurement can be found at the ICRC2015~\cite{icrc15_spectrum}.}, with an exposure that surpasses $30\,000\ {\rm km^2\,sr\,yr}$. The high energy region is dominated by SD events as the exposure is higher, while the lowest energy region benefits from the hybrid events. The exposure calculation for the  latter is performed using the periods of the telescopes operation and applying Monte Carlo simulations to obtain the aperture. All the measurements are combined into a single energy spectrum through a maximum likelihood fit procedure.

The final result is an energy spectrum with very small statistical uncertainty in most energy regions. Data has been fitted to a power law bellow the \emph{ankle} and a power law with a smoothed suppression above to emphasise the main features of the energy spectrum. Two features are clearly identified:  a change in the spectral index, around log(E/eV) $=18.72$, referred to as the \emph{ankle}; and a suppression of the flux above log(E/eV)$=19.63$. Although this suppression is consistent with the GKZ effect~\cite{GZK1,GZK2}, it could also be explained by the sources exhaustion to accelerate UHECRs. These two scenarios would have very different mass composition evolutions.  In fact, a large fraction of protons at the highest energy would be an indication of the presence of the GZK effect. On the other hand, a continuous evolution of the UHECRs towards a heavier composition would signal the existence of a source exhaustion mechanism. Thus, UHECRs mass composition is a key element to understand the sources mechanisms and UHECRs propagation.


\section{Primary mass composition}

The nature of UHECRs can be inferred from the shower observables. One of the most composition sensitive observables measured in Auger is the depth at which the shower reaches its maximum, $X_{\rm max}$. This quantity is measured by the Fluorescence Detector following quality cuts described in~\cite{XmaxMeas}. These cuts include reconstruction performance, cuts on the atmosphere conditions, namely the amount of aerosols, and cuts on the field of view so that the distribution does not get biased.

\begin{figure}[htbp]
	\centering %
	\begin{minipage}[c]{0.9\textwidth}%
		\centering \includegraphics[width=0.99\textwidth]{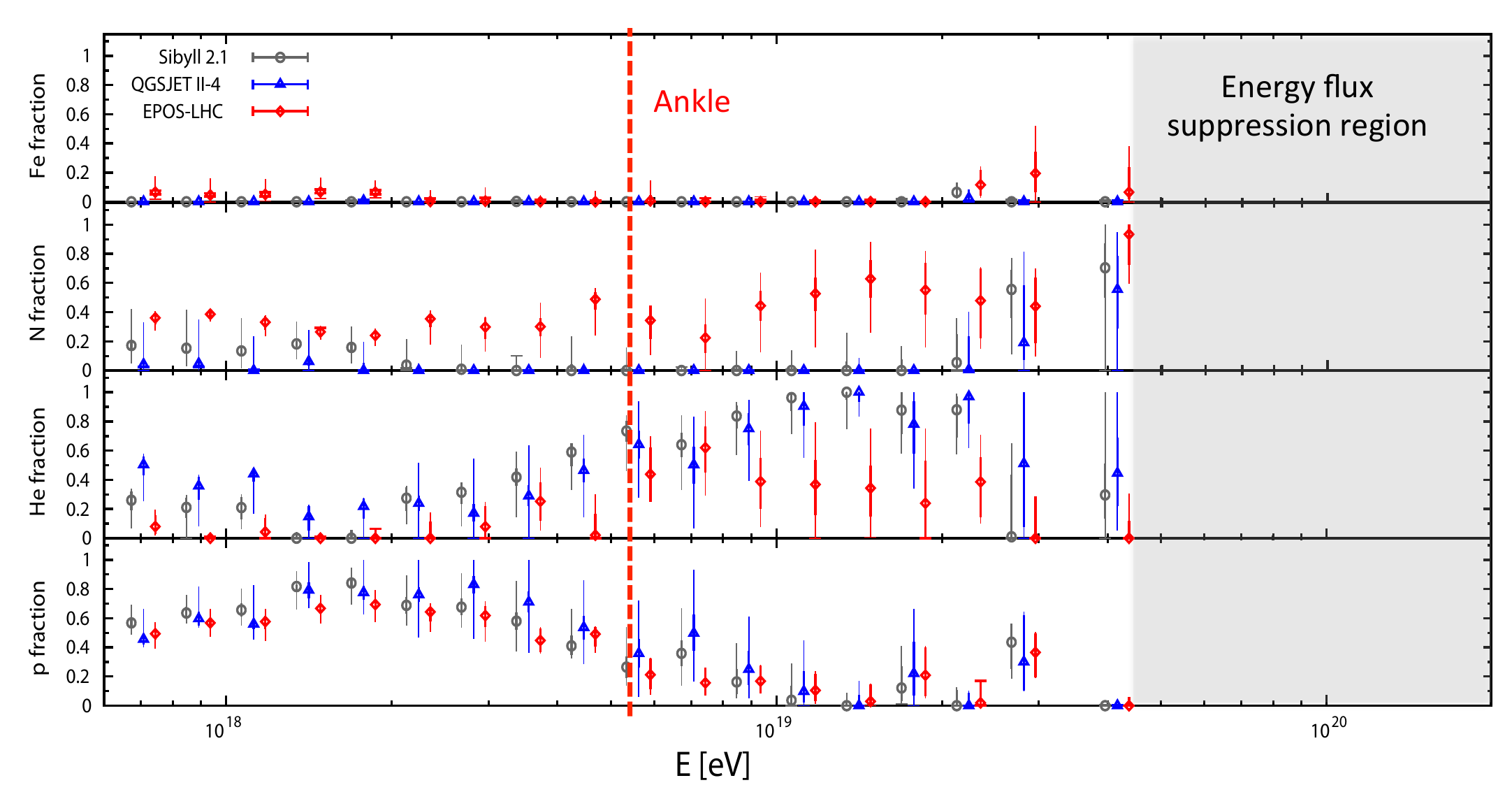} %
	\end{minipage}
	\protect\caption{Fitted fraction for the scenario of a complex mixture of protons, helium nuclei, nitrogen nuclei, and iron nuclei. The red dashed line indicates the position of the ankle with the shaded area the energy region where the flux rapidly decreases.~\cite{XmaxComp}}
\label{fig:XmaxComp}
\end{figure}

The primary mass composition was obtained by fitting the energy binned $X_{\rm max}$ distributions considering a specific hadronic interaction model. Four elements were considered for the primary mass composition: proton, helium, nitrogen and iron. The results are displayed in figure~\ref{fig:XmaxComp}. Although the fit results clearly depend on the choice of the hadronic interaction model general trends can be observed. At lower energies ($E\sim 10^{18}\,$eV) all models prefer a light composition with a significant fraction of protons. As the energy increases, the mass composition slowly evolves towards a heavier composition. At the highest possible FD energies, all the models prefer a composition dominated by intermediate mass states (helium and nitrogen) with almost no presence of proton nor iron. It is important to notice that, as this is FD data, there are no events for the energy region where the suppression of the energy spectrum occurs.

\section{Particle physics measurements}

The interpretation in terms of primary mass composition depends of the knowledge on the shower description. This means that the understanding of high-energy hadronic interactions is crucial. However, this is also an opportunity to do particle physics at energies that surpass those reached currently by the Large Hadron Collider (LHC). In this section we present some of the particle physics related measurements done at Auger.

\subsection{Proton-air cross-section}

\begin{figure}[htbp]
	\centering %
	\begin{minipage}[c]{0.45\textwidth}%
		\centering \includegraphics[width=0.99\textwidth]{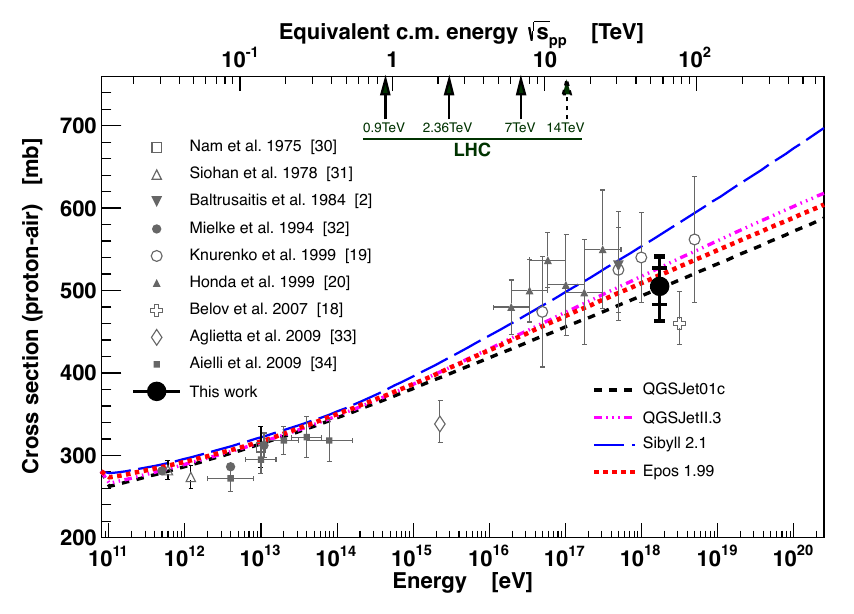} %
	\end{minipage}
	\begin{minipage}[c]{0.45\textwidth}%
		\centering \includegraphics[width=0.99\textwidth]{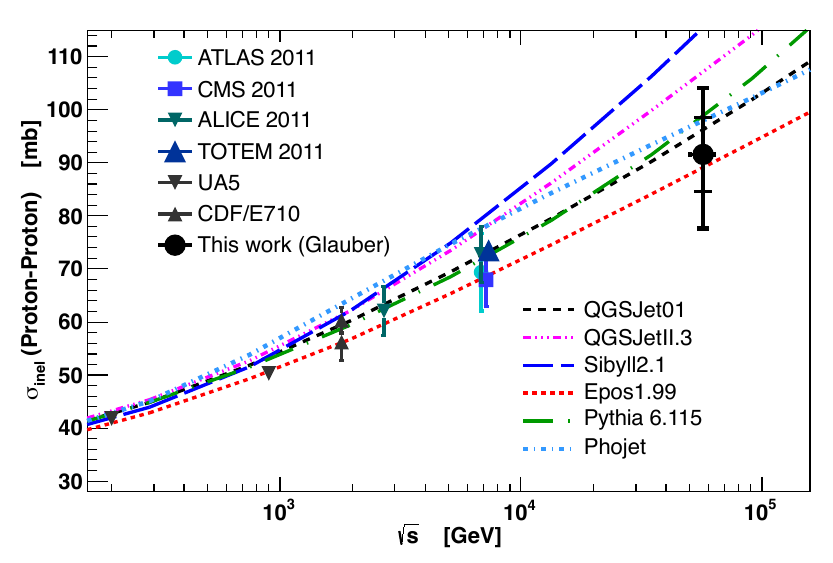} %
	\end{minipage}%
	\protect\caption{Left: Comparison between the Auger proton-air cross-section measurement and the result of previous experiments along with theoretical models predictions. The inner error bars are statistical, while the outer include systematic uncertainties for a helium fraction of 25\% and 10 mb for the systematic uncertainty attributed to the fraction of photons. Right: Comparison of derived proton-proton $\sigma_{\rm inel}$ to model predictions and accelerator data.\cite{AugerXsec}}
\label{fig:Xsec}
\end{figure}

The depth of the shower maximum, $X_{\rm max}$, depends on the shower development. In particular, the exponential tail of the $X_{\rm max}$ distribution is very sensitive to the primary cross-section. Moreover, among all the possible hadronic primaries, proton has the smallest cross-section and consequently, proton induced showers have the deepest $X_{\rm max}$. By selecting the deepest showers, if there is a sufficient amount of proton elements in data, it is possible to estimate the proton-air cross-section. Auger has performed this measurement using FD events with energy in the range $\log(E/eV) \in [18;18.5]$ (with an average value of $\sqrt{s}=57\,$TeV) and reported\footnote{an update of this measurement can be found at the ICRC2015~\cite{icrc15_Xsec}.} a proton-air cross-section of $\sigma_{\rm p-air}=[505\pm22\,{\rm (stat)}\,^{+28}_{-36}\,{\rm (syst)}]\,$mb (see figure~\ref{fig:Xsec} (left))~\cite{AugerXsec}. The main sources of systematic uncertainties for this measurement are the fraction of photons and helium. The first can be estimated from data while the latter was assumed to be less than $25\%$.

It is possible to interpret this measurement in terms of proton-proton inelastic cross-section using a Glauber formalism. This exercise is presented in figure~\ref{fig:Xsec} (right). It is interesting to note that the Auger data point is within $1\sigma$ of the accelerator measurements extrapolation.

\subsection{EAS muon content in inclined events}

\begin{figure}[htbp]
	\centering %
	\begin{minipage}[c]{0.45\textwidth}%
		\centering \includegraphics[width=0.99\textwidth]{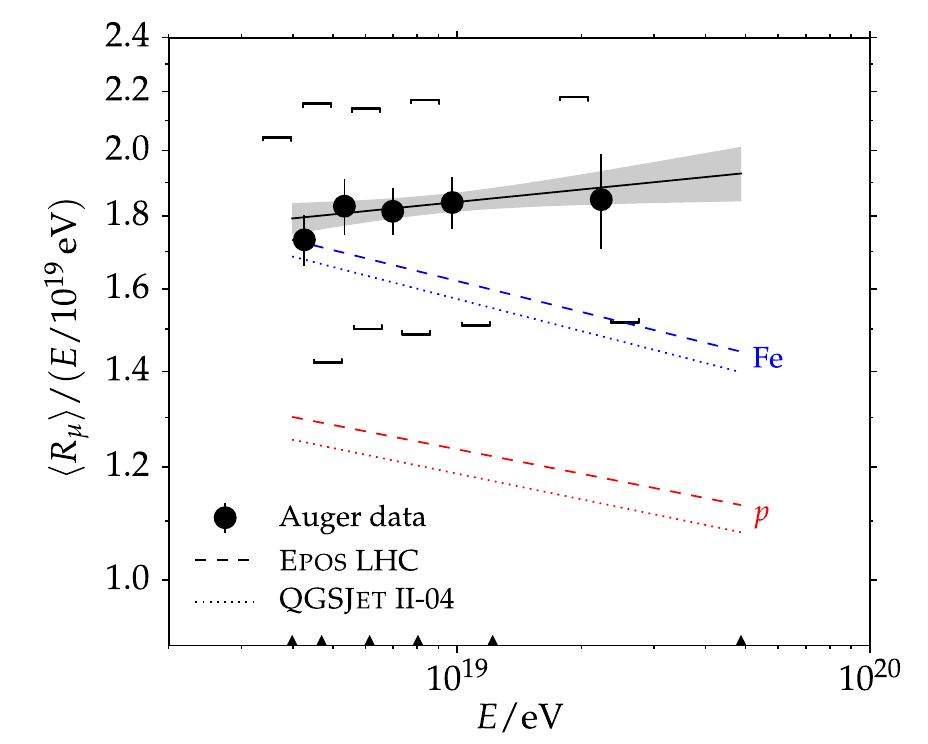} %
	\end{minipage}
	\begin{minipage}[c]{0.45\textwidth}%
		\centering \includegraphics[width=0.99\textwidth]{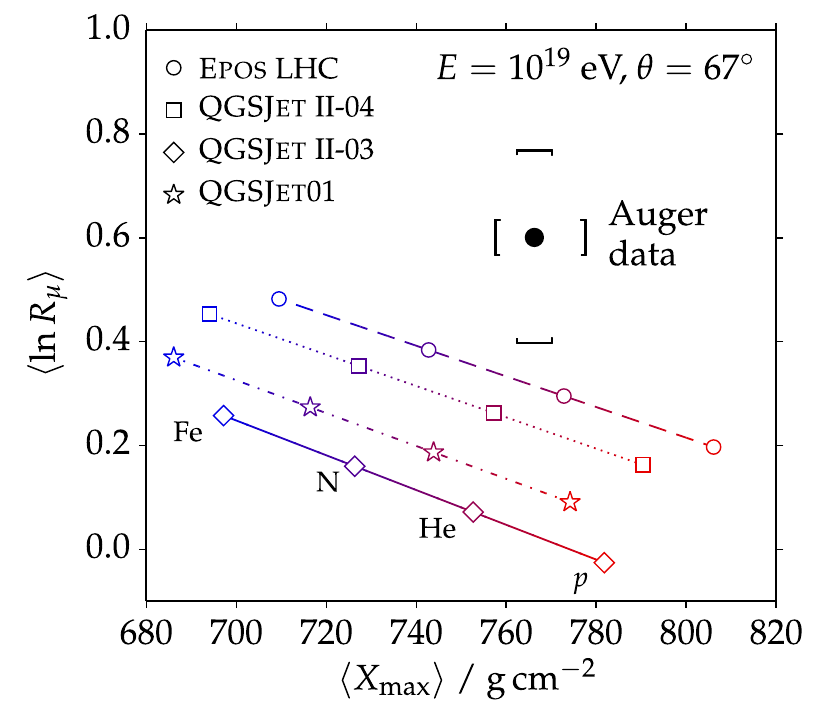} %
	\end{minipage}%
	\protect\caption{Left: Average shower muon content $\left< R_\mu \right>$ as a function of the shower energy $E$. Data is shown as the black dots together with its statistical uncertainty while the systematic measurement uncertainty is represented by the square brackets. The predictions of the models for proton and iron induced shower are shown as dashed lines. Right: Average logarithmic muon content as a function of the average shower depth at $10^{19}\,$eV. The predictions of the different hadronic interaction models under an assumption of a given primary mass composition are also shown.~\cite{MuonsInclined}}
\label{fig:muons}\label{fig:muonsXmax}
\end{figure}

Being produced in the decay of charged mesons and having a large probability of reaching the ground, muons are an important tool to assess the hadronic component of the shower. Although the SD is sensitive to both the electromagnetic and muonic shower components, in inclined showers most of the electromagnetic component of the shower gets absorbed in the atmosphere. Hence, the measurement of the signal at the ground allows to assess the EAS muon content. The results of this analysis as a function of energy are shown in figure~\ref{fig:muons} (left)~\cite{MuonsInclined}. Within the systematic uncertainties, data is compatible with a heavier composition in the full energy range. Moreover, the trend seems to indicate a transition in terms of mass composition, again towards heavier elements.

Interestingly, if one confronts the value of this measurement, $\left< R_\mu \right>$, with the $\left< X_{\rm max} \right>$ measurement at $E=10^{19}$ eV one finds that none of the present post-LHC tuned hadronic interaction models can provide a consistent solution in terms of primary mass composition. This conclusion is achieved even considering all the measurements systematic uncertainties.

\subsection{$X^\mu_{\rm max}$ vs $X_{\rm max}$}

Auger has also measured the profile of the depth at which muons are produced during the development of each shower, the muon production depth (MPD), and published the evolution of the average depth of the maximum of the MPD profile, $\left< X^\mu_{\rm max} \right>$, as a function of energy~\cite{MPD}. The measurement of the MPD profile is done with SD events using the geometrical delay of the muons, i.e. combining the shower geometry with the arrival time of these particles. The analysis needs to be performed in regions dominated by muons and as such it uses inclined events ($\theta \sim 60^\circ$) and regions away from the shower core. Similarly to the $X_{\rm max}$ results, but with larger uncertainties, the data indicates an increase of heavier elements as the energy increases. It is important to note that, as this analysis is performed with the SD, it can reach higher energies.

\begin{figure}[htbp]
	\centering %
	\begin{minipage}[c]{0.85\textwidth}%
		\centering \includegraphics[width=0.99\textwidth]{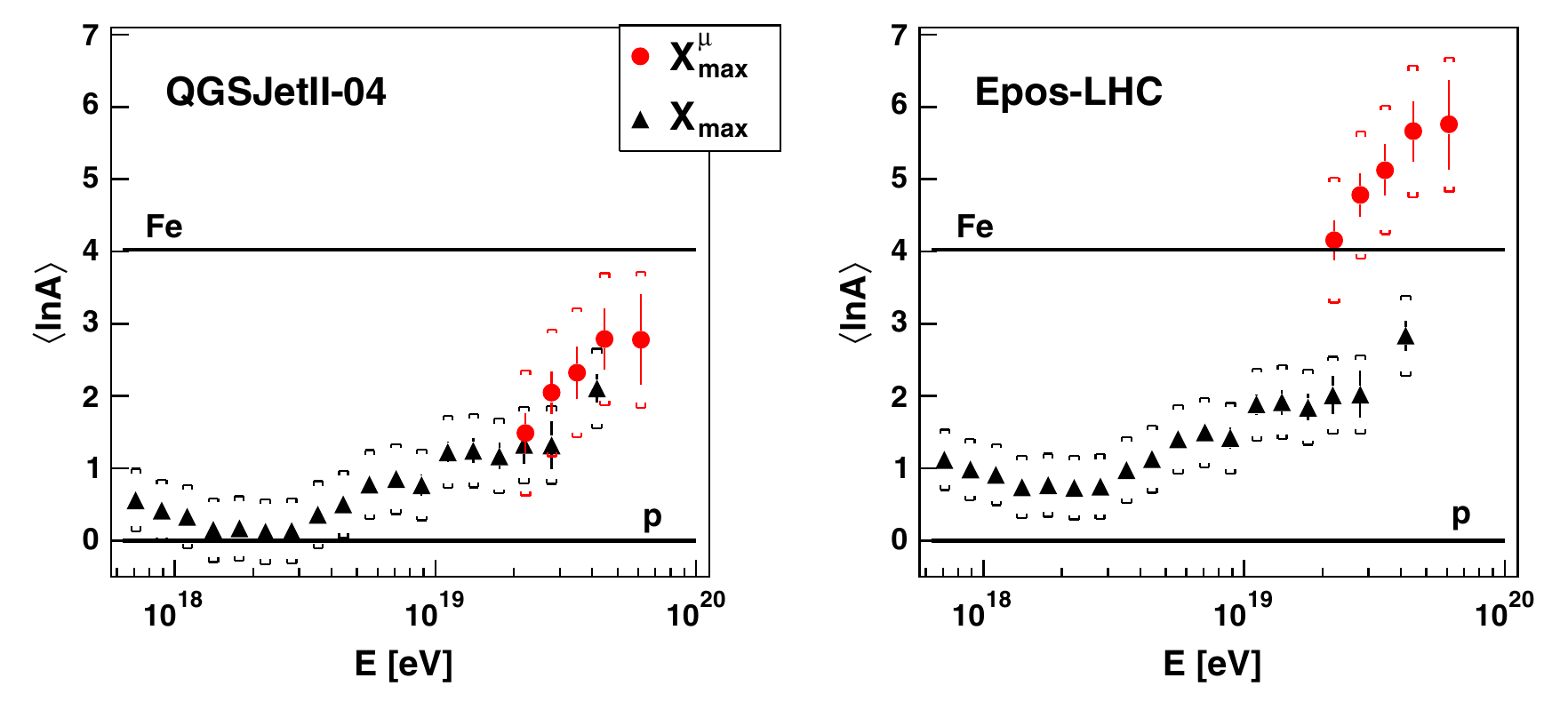} %
	\end{minipage}
	\protect\caption{Conversion of $\left< X^\mu_{\rm max} \right>$ (circles) and $\left< X_{\rm max} \right>$ (triangles) to $\left< \ln A \right>$ as a function of the shower energy. On the left (right) plot we use QGSJetII-04 (EPOS-LHC) as the reference hadronic interaction model. Brackets correspond to the systematic uncertainties.~\cite{MPD}}
\label{fig:XmumaxXmax}
\end{figure}

Similarly to what was presented in the previous section, the $\left< X^\mu_{\rm max} \right>$  results can be compared to the $\left< X_{\rm max} \right>$, measured by the FD, to test the shower description. In average, both shower observables should be interpreted in terms of the same primary mass composition. Failing to do so would reveal problems in the shower physics description. To perform this test both $\left< X_{\rm max} \right>$ and $\left< X^\mu_{\rm max} \right>$ are converted to the average logarithm of the primary mass number, $\left< \ln A \right>$, under the assumption of a given hadronic interaction model, figure~\ref{fig:XmumaxXmax}. From this figure it is clear that EPOS-LHC cannot provide a consistent solution in terms of mass composition for both observables. Notice also that although QGSJet-II.04 shows a consistent treatment of the shower considering $\left< X_{\rm max} \right>$ and $\left< X^\mu_{\rm max} \right>$, it fails to describe at the same time the EAS muon content, $\left< R_{\mu} \right>$, and $\left< X_{\rm max} \right>$.

\section{Final Remarks}

The Pierre Auger Observatory is the world's largest experiment dedicated to the study of UHECRs. It has measured the end of the cosmic ray energy spectrum with enough statistics to firmly establish the presence of an \emph{ankle} and a suppression at the highest energies. However, the interpretation of these structures in terms of astrophysical constraints depends on the nature of UHECRs.

Auger has presented measurements of shower observables sensitive to the primary mass composition. All the measurements indicate a lighter composition at lower energies, which becomes gradually heavier as the energy of the primary increases. The interpretation of data in terms of mass composition depends on the modelling of the shower. Auger has performed several measurements to assess the hadronic interaction properties and delivered some particle physics measurements. One of the most relevant is the measurement of the proton-air cross-section at $\sqrt{s}=57\,$TeV. Moreover, through the combination of several observables sensitive to both primary mass composition and hadronic interactions, Auger was able to demonstrate that none of the up-to-date hadronic interaction models is able to provide a fully consistent description of the shower. 

The obtained results pushed the collaboration to move towards an upgrade to measure the electromagnetic and muonic shower components at ground separately. This shall be achieved by installing a scintillator detector, which is mostly sensitive to the electromagnetic particles, on top of each water Cherenkov detector, which measures both components. Through this upgrade the observatory enhances not only the identification of the primary mass composition in a event-by-event basis, but also will be able to gain a deeper understanding of the shower physical mechanisms.

\bibliographystyle{JHEP}
\bibliography{bib_eps2015}

\end{document}